\documentclass[]{raa}  
\usepackage[flushleft]{threeparttable}
\usepackage{blindtext}
\usepackage{amsmath}
\usepackage{mathtools}
\usepackage{multirow}
\usepackage{comment}
\usepackage{enumerate}
\usepackage{enumitem}
\usepackage{natbib}
\usepackage{amssymb}

\let\oldAA\AA
\renewcommand{\AA}{\text{\normalfont\oldAA}}

\begin{document}

\title{Optical Flares in the Luminous Fast Blue Optical Transient AT2022tsd (``Tasmanian Devil")}

\volnopage{ {\bf 20XX} Vol.\ {\bf X} No. {\bf XX}, 000--000}
\setcounter{page}{1}

\author{Rachid Ouyed}
\institute{Department of Physics and Astronomy, University of Calgary, Calgary, AB, Canada; {\it rouyed@ucalgary.ca}}
 
\abstract{
We propose that luminous fast blue optical transients (LFBOTs) signal the
delayed conversion of a massive neutron star (NS; $M_{\rm NS}\gtrsim
M_{\rm NS,c}\sim1.8\,M_\odot$) into a highly magnetized hybrid star (HS) with $B_{\rm HS}\sim10^{15}$~G surface field; a QCD magnetar. 
This is the partial conversion channel in the Quark-Nona (QN) model where the core of the NS enters a quark phase
with spontaneous generation of extreme (i.e., up to $> 10^{18}$ G) magnetic field independent of the NS spin. 
  The process ejects $\sim10^{-2}\,M_\odot$ of the NS outermost layers
at $\sim0.1c$ (the QN ejecta) with a photon diffusion timescale of
a few days.  The powering of the QN ejecta by spin-down of a rapidly
 rotating HS (inherited from the parent NS) yields the LFBOT.
The fragmentation of the QN ejecta allows optical flares to arise from clumps
that become optically thin, releasing stored radiation energy (with
luminosities comparable to the LFBOT peak) on
light-crossing timescales of tens of minutes.
X-rays from the relativistic HS spin-down wind
escaping through optically thin gaps in the QN ejecta, and radio from
QN ejecta--medium interaction arise self-consistently from a single
physical engine. This framework reproduces key features of AT2022tsd, AT2020xnd,
AT2020mrf, and AT2018cow.
The neutron-rich, $r$-process-producing QN ejecta predicts kilonova-like
emission associated with LFBOTs in environments that do not host
neutron star mergers.
}

\authorrunning{Rachid Ouyed}
\titlerunning{Flares in AT2022tsd}
\maketitle

\section{Introduction}
\label{sec:introduction}

Luminous fast blue optical transients (LFBOTs) are a class of optical
transients characterized by rapid rise and decay times (on the order of days)
and high peak luminosities ($\gtrsim10^{43}$~erg~s$^{-1}$;
\citealt{drout_2014,pursiainen_2018,metzger_2022}).
The prototype event, AT2018cow
\citep{prentice_2018,rivera_2018,margutti_2019,perley_2019},
has been followed by several analogs, including AT2020xnd \citep{perley_2021},
AT2020mrf \citep{yao_2022}, AT2022tsd \citep{ho_2023a},
and AT2023fhn \citep{chrimes_2024a}.

Early-time spectra of LFBOTs are characterized by a hot, featureless blue
continuum with blackbody temperatures of $\sim(1.5$--$3)\times10^4$~K and a
notable absence of strong spectral lines, with hydrogen and helium features
emerging only at later times (e.g., \citealt{xiang_2021}).
Unlike core-collapse supernovae (CCSNe; \citealt{branch_2017}), LFBOTs evolve more rapidly and are
not powered by radioactive decay.
While their peak luminosities can rival those of superluminous supernovae
(SLSNe; e.g., \citealt{quimby_2013,galyam_2019} and references therein),
LFBOTs lack hallmark SLSN features such as broad O~II absorption and
nebular-phase spectra, and they arise in distinct environments
(e.g., \citealt{leloudas_2015,moriya_2024}).

LFBOTs are typically found in star-forming dwarf and spiral galaxies,
although some events, such as AT2023fhn, occur in intergalactic environments
while still exhibiting circumstellar properties similar to other members of
the class \citep{chrimes_2024a,chrimes_2024b}.
They do not strongly trace active star-forming regions within their hosts
and show a weaker dependence on metallicity than other extreme transients.
Their volumetric rate is estimated to be $\lesssim1\%$ of the CCSN rate
(e.g., \citealt{coppejans_2020,ho_2023b}).

Bright radio and X-ray emission indicate dense circumstellar interaction
and pre-explosion mass loss
\citep{ho_2019,huang_2019,coppejans_2020,bright_2022}.
A wide range of models has been proposed to explain these events, including
low-ejecta-mass CCSNe powered by central engines (e.g., magnetar spin-down
or black hole accretion; \citealt{margutti_2019,chen_2023}), shock interaction
\citep{fox_2019,pellegrino_2022}, embedded internal shocks
\citep{pasham_2021}, and tidal disruption events involving intermediate- or
stellar-mass black holes
\citep{kuin_2019,perley_2019,liu_2018,uno_2020b,kremer_2021}.
Alternative scenarios include jetted explosions in common-envelope systems
\citep{soker_2019,soker_2022} and fallback or electron-capture supernovae
\citep{lyutikov_2019,quataert_2019,piro_2020,uno_2020a,lyutikov_2022}.
Despite this diversity, no single model self-consistently explains the rapid
evolution, high luminosities, and multi-wavelength properties of LFBOTs.

In the QN model (see \citealt{ouyed_2022a,ouyed_2022b} for recent reviews),
a NS can undergo one of two conversion channels.
In the first, the NS undergoes a complete transformation into a quark star (QS),
relying on the hypothesis that absolutely stable strange quark matter exists
\citep{itoh_1970,bodmer_1971,terazawa_1979,witten_1984}.
In the second (we adopt here) only the innermost NS core undergoes
quark deconfinement, producing a strongly magnetized HS;
what we call the \emph{QCD magnetar}.
This partial conversion is comparatively more flexible, requiring only
deconfined up and down quarks and avoiding the more restrictive assumption
of stable strange quark matter.
We define $M_{\rm NS,c}$ as the critical NS mass above which the core density
$\rho_{\rm c}$ exceeds the quark deconfinement threshold
$\rho_{\rm dec}\sim2$--$5\,\rho_{\rm nuc}$ (e.g., \citealt{glendenning_1997,weber_2005,Alford2005} 
and references therein), where $\rho_{\rm nuc}$ is the nuclear saturation density.
For equations of state consistent with the $2\,M_\odot$ mass constraint
\citep{demorest_2010,antoniadis_2013} and multimessenger NS radius measurements
 ( e.g., \citealt{miller_2021}),
this condition is reached at $M_{\rm NS}> \sim 1.8\,M_\odot$ (e.g., \citealt{annala_2018,baym_2018} and references therein).
 Hereafter, we adopt $M_{\rm NS, c} = 1.8M_{\odot}$, while noting that the exact value remains uncertain.
 
Three consequences of this partial transition in the QN model are relevant to the present paper:
(i) certain quark-matter phases at these densities can sustain internal
magnetic fields up to $\sim10^{18}$~G
\citep[e.g.,][]{iwazaki_2005,ebert_2005,dvornikov_2016,efrain_2021},
translating to a surface field $B_{\rm HS}\sim10^{15}$~G under a dipolar
configuration; (ii) $\sim10^{-2}\,M_\odot$ of  outer NS layers is expelled
at $v\sim0.1c$ with kinetic energy $\sim10^{50}$~erg
(\citealt{keranen_2005}); the QN ejecta's photon diffusion timescales is order of a few days only; (iii) The neutron-rich
QN ejecta  is shown to be favorable to r-process nucleosynthesis \citep{jaikumar_2007}. 

In \citet{ouyed_2025,ouyed_2026}, we proposed that LFBOTs arise from this
partial NS-to-HS transition when two additional conditions are met:
(i) the NS is born rapidly rotating, with a spin period of a few milliseconds;
(ii) the QN occurs before the NS has spun down appreciably,
i.e., at $t_{\rm QN}<t_{\rm NS,SpD}$, where $t_{\rm NS,SpD}$ is the NS
spin-down timescale prior to the QN.
These conditions ensure that the QCD magnetar inherits rapid rotation from
its progenitor NS and can efficiently power the QN ejecta via spin-down,
producing a transient with peak luminosity, rise and decay times consistent with LFBOTs.

Whether the resulting LFBOT is directly observable depends on the
environment in which the QN occurs.
If the NS approaches $M_{\rm NS,c}$ through accretion from a binary
companion, the QN occurs in isolation 
long after the SN ejecta has dissipated and the LFBOT is directly observable.
If the QN follows a CCSN while the SN ejecta is still dense and optically
thick, the LFBOT emission is reprocessed, yielding a SLSN \citep{ouyed_2026}.
In the CCSN case, a directly visible LFBOT results only if the 
QN occurs after the SN had dissipated. In this case,  the location of such events relative to star-forming regions depends on the NS 
progenitor's evolutionary history

In this paper, we extend the QN framework for directly observable LFBOTs
to show that the fragmentation of the QN ejecta (which is dynamically different from that of NS mergers) into clumps that become optically
thin on timescales of minutes naturally produces rapid optical flares with
luminosities comparable to the LFBOT peak.
X-ray emission arises from spin-down power escaping through low-optical-depth
regions in the fragmented ejecta, while radio emission is generated by
interaction of the QN ejecta with the ambient medium.
We demonstrate that this model simultaneously and self-consistently reproduces
the optical, X-ray, and radio properties of AT2022tsd and other LFBOTs.

This paper is organized as follows.
In \S~\ref{sec:ourmodel}, we describe the QN model and the NS-to-HS conversion.
In \S~\ref{sec:emission}, we present the resulting optical, X-ray, and radio
emission and apply the model to AT2022tsd and related events.
In \S~\ref{sec:discussion}, we discuss the implications and limitations of
the model before summarizing our conclusions in \S~\ref{sec:conclusion}.

\section{Our model}
\label{sec:ourmodel}

Hereafter, dimensionless quantities are defined as $f_x = f/10^x$ in cgs
units unless stated otherwise. We work in the NS rest frame.

\subsection{The NS-to-HS Conversion}
\label{sec:NS-to-HS}

We consider a NS with mass $M_{\rm NS}\gtrsim M_{\rm NS, c}\sim 1.8M_{\odot}$, radius $R_{\rm NS}=12$~km,
birth period $P_{\rm NS}=5$~ms ($10^{-2.3}$~s in our equations),
and surface magnetic field $B_{\rm NS}\sim10^{12.5}$~G. The spin-down formalism adopted
here for both the NS and HS follows
\citet{michel_1970,manchester_1977,shapiro_1983,lyne_1993,contopoulos_1999,%
spitkovsky_2006,lorimer_2004,lattimer_2005}. We assume an aligned, force-free magnetized wind with moment of inertia
$I_{\rm NS}\simeq2\times10^{45}$~g~cm$^2$.  The NS rotational energy is
\begin{equation}
\label{eq:ENSrot}
E_{\rm NS,rot} \simeq 1.2\times10^{51}\,P_{\rm NS,-2.3}^{-2}\ {\rm erg},
\end{equation}
and the characteristic  spin-down timescale is
\begin{equation}
\label{eq:tNSSpD}
t_{\rm NS,SpD} \simeq 39.5\ {\rm yr}\ B_{\rm NS,12.5}^{-2}\,P_{\rm NS,-2.3}^{2} \ .
\end{equation}
This is the characteristic timescale on which the NS loses an $e$-folding
of its initial spin-down luminosity due to magnetic dipole radiation.

At time $t_{\rm QN}$, the NS undergoes a phase transition to a HS.
The remaining rotational energy fraction is
$(1+t_{\rm QN}/t_{\rm NS,SpD})^{-2/(n-1)}$. 
Here, $n$ is the braking index is defined as $
n = \frac{\Omega \, \ddot{\Omega}}{(\dot{\Omega})^2}$ (e.g. \citealt{shapiro_1983,lorimer_2004}),
and it characterizes the torque mechanism responsible for the spin-down of a NS, determining how the rotational frequency $\Omega$ decreases over time.
We adopt a braking index $n=3$, appropriate for magnetic dipole radiation,
where the NS spin-down torque scales as $\dot{\Omega}\propto-\Omega^3$.
This yields the period evolution $P(t)=P_0(1+t/t_{\rm NS, SpD})^{1/2}$,
from which the HS inherits the spin period
\begin{equation}
\label{eq:PHS}
P_{\rm HS} = P_{\rm NS}\left(1+\frac{t_{\rm QN}}{t_{\rm NS,SpD}}\right)^{1/2}\ .
\end{equation}
I.e. the HS rotational energy at birth is $E_{\rm HS, rot}= \frac{1}{2}I_{\rm HS}(2\pi /P_{\rm HS})^2$
 where $I_{\rm HS}=I_{\rm NS}$.

For the LFBOT to be powered efficiently, the HS must inherit rapid rotation,
requiring $t_{\rm QN}<t_{\rm NS,SpD}$, i.e.,
\begin{equation}
\label{eq:tQNmax}
t_{\rm QN} < 39.5\ {\rm yr}\ P_{\rm NS,-2.3}^{2}B_{\rm NS,12.5}^{-2}.
\end{equation}
We can define a critical NS magnetic field $B_{\rm NS,c}\sim10^{13.5}$~G such that
for $B_{\rm NS}>B_{\rm NS,c}$, the constraint becomes
$t_{\rm QN}\lesssim150$~days. I.e., for the CCSN channel, even if the condition for the LFBOT formation is satisfied
 the emission will be buried within the dense and optically thick SN ejecta resulting in a 
SLSN instead \citep{ouyed_2026}.  

Fits to double-peaked SLSNe \citep[e.g.,][]{ouyed_2026} suggest $t_{\rm QN}$ values of weeks to months. In this interpretation, the first peak arises from spin-down powering of the SN ejecta during the NS phase, followed by a second phase of spin-down associated with the QCD magnetar (i.e., the HS). The relative strengths of the two peaks depend on the fraction of rotational energy inherited by the HS at $t_{\rm QN}$.
These fits imply that $t_{\rm QN}$, set by the quark nucleation timescale in the deconfined core of massive NSs, occurs on timescales shorter than those required for typical SN ejecta to become optically thin (i.e., $\lesssim$ a few years).
If this behavior is universal, it would imply that naked LFBOTs are more likely associated with accreting NSs that reach $M_{\rm NS,c}$ well after birth.  Nevertheless, as noted earlier, under special circumstances a naked LFBOT may still be linked to a CCSN; for example, if the progenitor NS had traveled far from star-forming regions prior to the explosion.

Following the transition, the same spin-down formalism applies to the HS
with $B_{\rm NS}\to B_{\rm HS}$ and $P_{\rm NS}\to P_{\rm HS}$.
The HS retains the period $P_{\rm HS}$ from Eq.~(\ref{eq:PHS}) but acquires
a significantly stronger magnetic field, $B_{\rm HS}\sim10^{15}$~G.
Its characteristic spin-down timescale is
\begin{equation}
\label{eq:tHSspd}
t_{\rm HS,SpD} \simeq 0.15\ {\rm days}\ B_{\rm HS,15}^{-2}\,P_{\rm HS,-2.3}^{2},
\end{equation}
and its spin-down luminosity is
\begin{equation}
\label{eq:LHSSpD}
L_{\rm HS,SpD} \sim 9.6\times10^{46}\ {\rm erg\ s}^{-1}\
B_{\rm HS,15}^{2}\,P_{\rm HS,-2.3}^{-4}
\left(1+\frac{t}{t_{\rm HS,SpD}}\right)^{-2}.
\end{equation}
This energy is injected behind the QN ejecta, powering the LFBOT phase
\citep{ouyed_2025,ouyed_2026}.

The spin-down luminosity $L_{\rm HS,SpD}$ is carried by a relativistic
magnetized wind of $e^\pm$ pairs, directly analogous to the pulsar wind driven
by NS spin-down. This wind is thermalized at its termination shock inside the
QN ejecta (as in SN ejecta case; \citealt{ostriker_1971,kasen_2010,woosley_2010}), powering the LFBOT. 
Crucially, the same mechanism also generates
synchrotron and inverse Compton radiation at X-ray energies.
The X-ray emission in our model is therefore not an additional assumption but
a direct and natural consequence of the same spin-down wind physics that powers
the LFBOT; its luminosity and temporal evolution are discussed in
\S~\ref{sec:xray}.

Considering the regime where diffusion regulates the buildup of energy, and directly solving the full energy equation, gives us an estimate of the total  rotational energy deposited  up to time  as  $E_{\rm HS, SpD}(t)\sim E_{\rm HS, rot}\times t/(t+t_{\rm QN,d})$. 
 By equating the blackbody (BB) luminosity to the spin-down energy injection at time $t$ 
from the HS, we write
\begin{equation}
4\pi R_{\rm QN}^2 \sigma_{\rm SB} T_{\rm BB}^4 \sim \frac{E_{\rm HS, rot}}{t_{\rm QN,d}} \times \frac{t}{t + t_{\rm QN,d}},
\end{equation}
where $R_{\rm QN} = v_{\rm QN} t$ and $\sigma_{\rm SB}$ is the
Stefan--Boltzmann constant. Solving for the effective temperature yields
\begin{equation}
\label{eq:TBB}
T_{\rm BB}(t) \sim 5.8\times10^4~{\rm K}~P_{\rm HS,-2.3}^{-1/2}
\kappa_{\rm QN,1}^{-1/2} M_{\rm QN,-2}^{1/16} E_{\rm QN,50}^{-3/16}
\times \left[ \frac{({\rm day}/t)^2}{(1 + t_{\rm QN,d}/t)} \right]^{1/4}\ .
\end{equation}
Because of the ejecta's heavy composition (\citealt{jaikumar_2007}),  
its optical opacity's fiducial value we take be $\kappa_{\rm QN}= 10$ cm$^2$ gm$^{-1}$ (\citealt{kasen_2013}). 
 
\subsection{QN ejecta fragmentation}
\label{sec:fragmentation}

Physically, the QN ejecta consists of the NS outer layers accelerated by
the fireball generated in the quark--hadron phase transition.
The energy is initially deposited as heat via neutrinos, with the resulting
fireball acting like a piston that pushes the overlying layers outward while
expanding adiabatically \citep{keranen_2005,ouyed_leahy_2009}.
Most of the energy is converted into bulk kinetic motion rather than heating,
so the dense ejecta avoids strong shock heating.
Only the extremely low-density outer layers experience modest shocks;
their mass is negligible compared to the total ejecta mass.
This inside-out, mechanically driven ejection is fundamentally different
from the outside-in tidal stripping that characterizes NS merger ejecta,
where shocks rapidly homogenize the material.
Here, the NS outer layers are cold and partially crystalline at the moment
of ejection, and the absence of strong shock heating allows the ejecta to
retain the pre-existing density inhomogeneities of the NS crust during
early expansion. These inhomogeneities, we speculate, seed the fragmentation that occurs later.

The QN ejecta propagates at a speed 
\begin{equation}
v_{\rm QN} = \sqrt{2 E_{\rm QN}/M_{\rm QN}} \sim 0.1c\, E_{\rm QN,50}^{1/2} M_{\rm QN,-2}^{-1/2},
\end{equation}
until it reaches the deceleration radius, 
\begin{equation}
\label{eq:R-dec}
R_{\rm dec} = \left(\frac{3 M_{\rm QN}}{4\pi n_{\rm amb} m_{\rm p}}\right)^{1/3},
\end{equation}
where it sweeps up an ambient mass comparable to its own. Here $n_{\rm amb}$ is the
 ambient number density and $m_{\rm p}$ the proton mass. From $R_{\rm dec} = v_{\rm QN} t_{\rm dec}$, the deceleration timescale is
\begin{equation}
\label{eq:tdec}
t_{\rm dec} \simeq 52.2\ {\rm days} \frac{M_{\rm QN,-2}^{5/6}}{E_{\rm QN,50}^{1/2} n_{\rm amb,6}^{1/3}}.
\end{equation}

The density of the ejecta at the deceleration radius can be estimated as
\begin{equation}
\rho_{\rm QN,dec} = \frac{\rho_{\rm amb}}{3} \frac{R_{\rm dec}}{\Delta R_{\rm dec}},
\end{equation}
where we made use of $M_{\rm QN}= \rho_{\rm QN,dec}  4\pi R_{\rm dec}^2 \Delta R_{\rm dec}$ and of Eq. (\ref{eq:R-dec})
 with $\Delta R_{\rm dec} = 10^{-2} R_{\rm dec}$. 
Hereafter, we assume a thin shell with $\Delta R_{\rm QN} = 10^{-2} R_{\rm QN}$ with $R_{\rm QN}=v_{\rm QN}t$. 

The QN ejecta's diffusion timescale is $t_{\rm QN}= \sqrt{2\kappa_{\rm QN} M_{\rm QN}/\beta c v_{\rm QN}}$ 
with $\beta=4\pi^3/9$  a geometric correction factor (\citealt{arnett_1982}).   This gives

\begin{equation}
\label{eq:TQNd}
t_{\rm QN, d}\sim 6.3\ {\rm days}~ \kappa_{\rm QN, 1}^{1/2} M_{\rm QN, -2}^{3/4} E_{\rm QN, 50}^{-1/4}\ ,
\end{equation}
where we made use of $v_{\rm QN}= \sqrt{2 E_{\rm QN}/M_{\rm QN}}$. 

As the ejecta expands and is decelerated by the ambient medium,
radiation pressure support within the ejecta drops.
 Between the diffusion timescale $t_{\rm QN,d}$ and the transparency
timescale $t_{\rm QN,tr}$, the declining radiation pressure can no longer
smooth out the pre-existing density contrasts, and the mechanically coherent
but inhomogeneous ejecta (acquired during the QN event proper) begins to fragment into distinct clumps.
Fragmentation is therefore a two-stage process: the initial cold,
structured ejection seeds the density contrasts, and the subsequent loss
of radiation pressure support during deceleration allows those contrasts
to develop into optically distinct fragments.
Other processes, such as Rayleigh--Taylor instabilities at the
ejecta--ambient interface or anisotropies in the NS crust structure,
could further contribute to clumpiness, but the primary driver n our model in its current form is the
cold, inhomogeneous nature of the QN ejecta at launch.

\section{Emission}
\label{sec:emission}

\subsection{Optical: the LFBOT}
\label{sec:optical}

The total  rotational energy deposited  up to time $t$ is  $E_{\rm HS, SpD}\sim E_{\rm HS, SpD}(0)\times t/(t+t_{\rm QN,d})$ and since 
$t_{\rm HS, SpD}< t_{\rm QN, d}$, the LFBOT luminosity  peaks
at around $t_{\rm QN, d}$  with $L_{\rm LFBOT, pk} \sim E_{\rm HS, SpD}/2t_{\rm QN, d}$ (up to factors of order unity depending on diffusion details; 
 e.g. \citealt{kasen_2010}). 
Hereafter, the subscript ``pk" stands for peak. We find

\begin{equation}
\label{eq:Llfbot}
L_{\rm LFBOT, pk} \sim 1.9\times 10^{45}\ {\rm erg\ s}^{-1}\times \frac{E_{\rm QN, 50}^{1/4}}{P_{\rm HS, -2.3}^2 \kappa_{\rm QN, 1}^{1/2} M_{\rm QN, -2}^{3/4} }\ .
\end{equation}

The average time it takes the QN to become optically thin (i.e. $\tau_{\rm QN}\rho_{\rm QN}\Delta R_{\rm QN}\sim 1$) is
\begin{equation}
\label{eq:tQNtr}
t_{\rm QN, tr}\sim 15\ {\rm days}~ \kappa_{\rm QN, 1}^{1/2} M_{\rm QN, -2} E_{\rm QN, 50}^{-1/2}\ ,
\end{equation}
with $M_{\rm QN}= 4\pi R_{\rm QN}^2 \rho_{\rm QN} \Delta R_{\rm QN}$ and $R_{\rm QN}= v_{\rm QN}t$. 
 Since $v_{\rm QN}< 0.5c$ (\citealt{keranen_2005}), we have  $t_{\rm QN, tr}> t_{\rm QN, d}$ indicating that diffusion is nearly complete before the ejecta becomes transparent; $t_{\rm QN, tr}$  is an average timescale, and we expect transparency to be reached in a spatially non-uniform manner across the ejecta.

\begin{figure*}[t!]
\begin{center}
\includegraphics[scale=0.4]{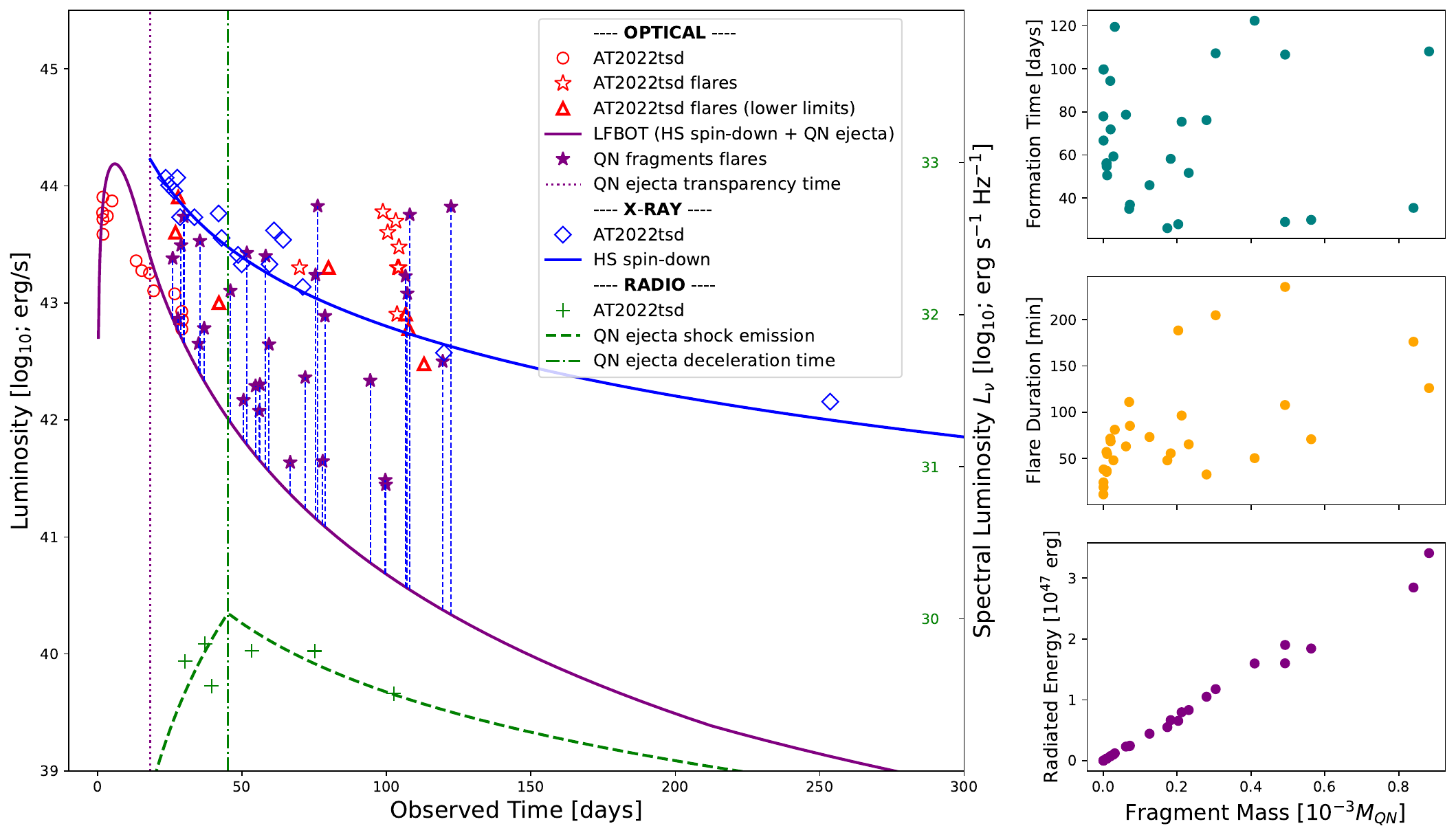}
\caption{Our model fits to the optical (bolometric), X-ray, and radio light curves of AT2022tsd are shown in the left panel. The optical (bolometric) and X-ray luminosities are read from the left y-axis; the radio spectral luminosity at the observed frequency listed in Table~1 is read from the \emph{right} y-axis. Solid colored curves show the model fits. The \emph{dotted vertical line} marks $t_{\rm QN,tr}$ (Eq.~\ref{eq:tQNtr}), the time when the mean QN ejecta becomes optically thin. The \emph{dashed-dotted vertical line} marks $t_{\rm dec}$ (Eq.~\ref{eq:tdec}), the ejecta deceleration time. Optical flares (short-duration spikes) originate from optically thin fragments that release their stored radiation energy on their light-crossing timescale (see \S\,\ref{sec:optical-flares}). The thin dashed blue vertical lines indicate the luminosity jump as the fragments (emitting at the LFBOT level) transition from optically thick to optically thin, releasing stored radiation energy. The right panels show fragment properties as functions of fragment mass $m_{\rm f}$ (in units of $10^{-3}\,M_{\rm QN}$), ordered from smallest to largest: formation time $t_{\rm f}$ (top), duration $\Delta t_{\rm f}$ (middle), and radiated energy $E_{\rm f,rad}$ (bottom). Data adapted from \citet{ho_2023a}.  All sampled fragments satisfy the $\tau_{\rm f}<1$ constraint.}
\label{fig:LC-tsd}
\end{center}
\end{figure*}

We follow \citet{chatzopoulos_2012} when computing the spin-down powered light-curves 
and include leakage of energy via hard emission with a leakage parameter $A_{\rm HS}= 9 \kappa_{\rm QN}M_{\rm QN}^2/40\pi E_{\rm QN}$ (\citealt{wang_2015}).
 Our model shows good agreement with the optical LC of AT2022tsd as shown in Figure \ref{fig:LC-tsd} 
 with the corresponding parameters listed in Table \ref{table:parameter-fits}. 
 
 %
 %
\begin{table*}[t!]
\begin{center}
\caption{Parameters for the model fits.$^\dagger$}
 \label{table:parameter-fits}
   \begin{tabular}{|c|c|c||c||c|c|c|c|}\hline
   Source & \multicolumn{2}{|c||}{LFBOT$^{\dagger\dagger}$} &  \multicolumn{1}{|c||}{Flares}  &  \multicolumn{4}{|c|}{Radio} \\\hline
   & $P_{\rm HS}$ (ms) & $B_{\rm HS}$ (G) & $\alpha_{\rm f}=t_{\rm f, pk}/t_{\rm QN, tr}$  & $n_{\rm amb}$ (cm$^{-3}$) & $\epsilon_{\rm e}$ &  $p$ & $\nu_{\rm obs}$ (GHz) \\\hline
  AT2022tsd & 8.5 & $6\times 10^{14}$ & 1.0&  $3\times 10^6$ & 0.073 &  2.4 & 100.0 \\\hline
  AT2020xnd & 9.0 & $9\times 10^{14}$  & 0.3&  $5\times 10^6$ & 0.08 &  2.4 & 100.0\\\hline
   AT2020mrf & 11.5 & $5.5\times 10^{14}$ & 0.3& $3\times 10^6$ & 0.03  & 2.4 & 10.0 \\\hline  
   AT2018cow & 6.0 & $8\times 10^{14}$  & 0.3& $2\times 10^7$ & 0.1  & 2.9 & 230.0\\\hline
  \end{tabular}\\
  \end{center}
  $^\dagger$ Kept fixed are: $E_{\rm QN}= 10^{50}$ erg, $M_{\rm QN} = 10^{-2}M_{\odot}$, $\kappa_{\rm QN}=10$ cm$^2$ gm$^{-1}$,
   $k_{\rm time}=k_{\rm radius}=1.5$, $R_{\rm f, pk}=6\times 10^{13}$ cm and $\epsilon_{\rm B}=0.01$.\\
  $^{\dagger\dagger}$ Also used to fit X-ray emission with $L_{\rm x}\sim L_{\rm HS, SpD}$ (see \S~ \ref{sec:xray}).
  \end{table*}
 %
 %

 \begin{figure}[t!]
\begin{center}
\includegraphics[scale=0.4]{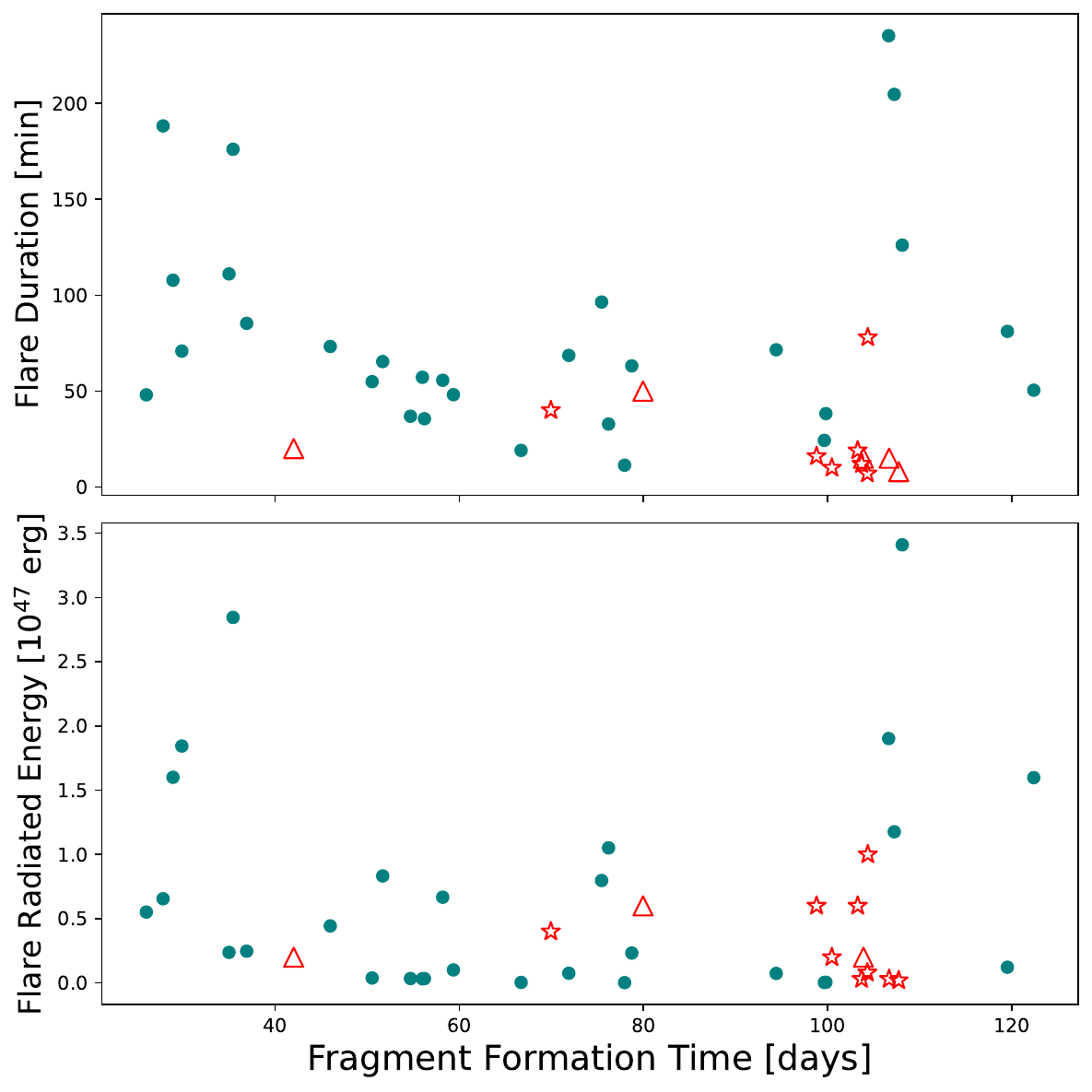}
\caption{Comparison of modeled flare durations (top panel) and radiated energies (bottom panel) as functions of fragment formation time. Modeled values are shown as solid circles, observed values as open symbols, and open triangles indicate lower limits. The data, here and in the subsequent figures, are adapted from \citet{ho_2023a}.}
\label{fig:compae-to-Ho}
\end{center}
\end{figure}

 \begin{figure}[t!]
\centering
\includegraphics[scale=0.4]{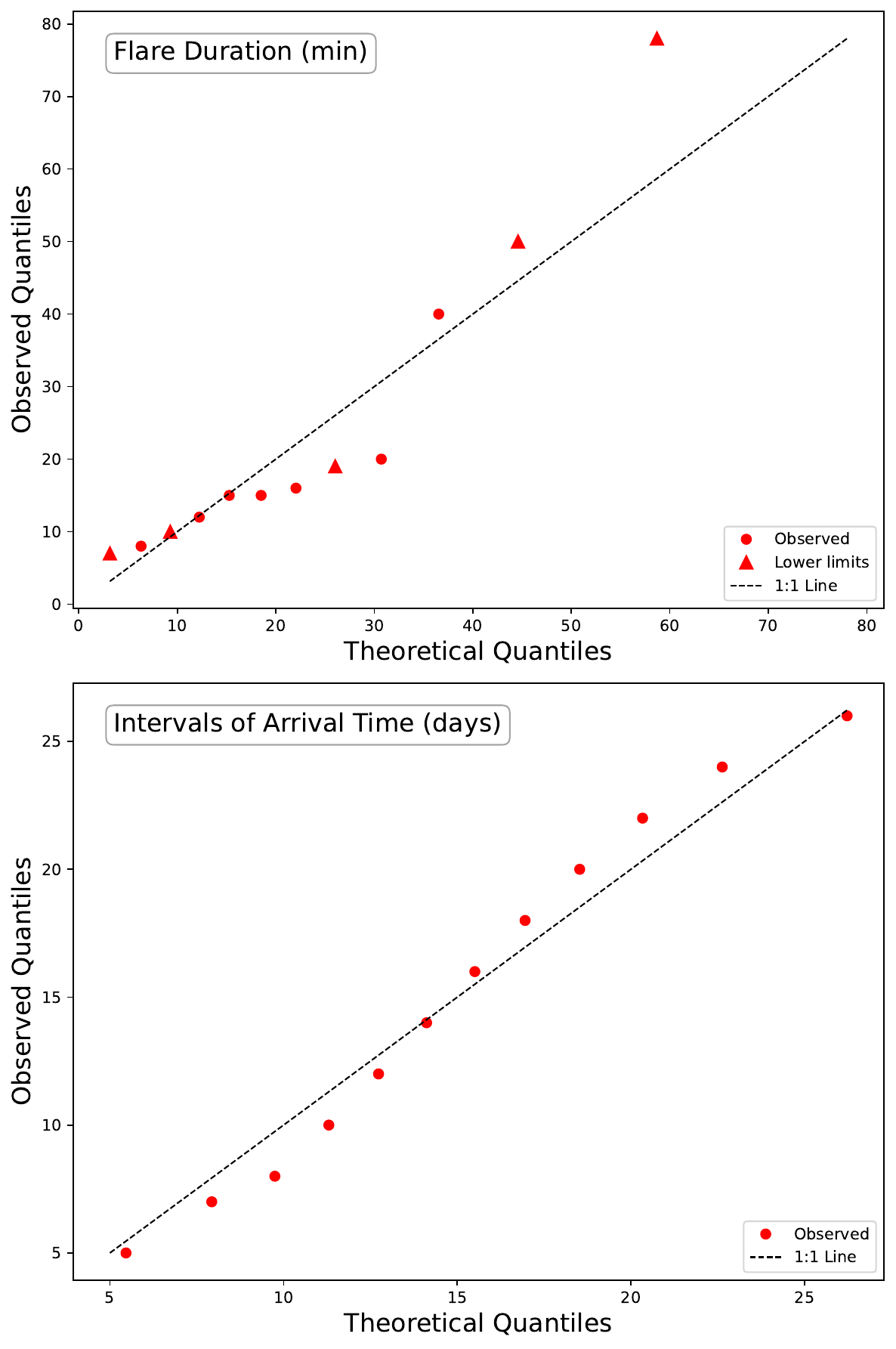}
\caption{Data quantiles plotted against theoretical quantiles from the WPD and WFTD distributions (our model). The black dashed line represents the 45$^\circ$  reference line, indicating a perfect fit.}
\label{fig:Q-Qplot}
\end{figure}

%
 %
 
 \subsection{Optical: the flares}
\label{sec:optical-flares}

We define $R_{\rm f}$ and $t_{\rm f}$   as the QN ejecta's 
 fragments'  characteristic  radius and formation time (since the NS conversion), respectively. 
 
To model the fragments' radius distribution, we employed a stochastic sampling approach based on the Weibull probability distribution (WPD; \citealt{weibull_1951}).
Fragment radii were sampled iteratively from a WPD  with the scale parameter $\lambda_{\rm radius}= R_{\rm f, pk}(\frac{k_{\rm radius}}{k_{\rm radius}-1})^{1/k_{\rm radius}}$ defining the most probable fragment radius and $R_{\rm f, pk}$ the peak fragment size; $k_{\rm radius}>1$ is the shape parameter. 
WPD is selected due to its flexibility in representing a range of fragment size distributions, particularly its ability to capture both broad ($k_{\rm radius}<2$) and sharply peaked ($k_{\rm radius}>2$) radius spectra. 

The fragment formation time $t_{\rm f}$ is sampled using a Weibull failure time model (WFTD) where fragments detach from the main QN ejecta at different times. 
The scale parameter is set so that the characteristic fragment formation time is a fraction of the QN ejecta transparency timescale $t_{\rm f, pk}=\alpha_{\rm f}t_{\rm QN, tr}$ with $\alpha_{\rm f}\le 1$ and $t_{\rm QN, tr}$ given by Eq. (\ref{eq:tQNtr}); i.e. $\lambda_{\rm time} = t_{\rm f, pk}(\frac{k_{\rm time}}{k_{\rm time}-1})^{1/k_{\rm time}}$.
  The distribution shape  can vary from a spread in fragmentation time ($k_{\rm time} < 2$) 
 to clustering of fragmentation events  around  $ t_{\rm f, pk}$ ($k_{\rm time} >2$).
 The total number of fragments was not fixed in advance. Instead, fragment masses ($m_{\rm f}\propto \rho_{\rm f}R_{\rm f}^3$) were added sequentially  and accumulated while
 ensuring that  their combined mass does not exceed $M_{\rm QN}$.

At time $t_{\rm f}$, the fragment density ($\rho_{\rm f}$) and temperature ($T_{\rm f}=T_{\rm BB}(t_{\rm f})$) are set to those of the QN ejecta.  I.e. 
 $\rho_{\rm f}=\rho_{\rm QN, f}= M_{\rm QN}/4\pi R_{\rm QN, f}^2\Delta R_{\rm QN, f} = (R_{\rm QN}/\Delta R_{\rm QN}) M_{\rm QN}/4\pi R_{\rm QN, f}^3$ 
   with the size of the QN ejecta then being $R_{\rm QN, f}= v_{\rm QN} t_{\rm f}$ so that $\rho_{\rm f}\propto t_{\rm f}^{-3}$.   
    The fragment temperature  is set by the HS spin-down energy
injected into the QN ejecta prior to $t_{\rm f}$, not by adiabatic free
expansion from the NS surface (see Eqs.~(\ref{eq:LHSSpD}) and (\ref{eq:TBB})). E.g., 
 fragments detaching at $t_{\rm f}\sim6$--20~days
therefore carry $T_{\rm f}\sim(1.5$--$2.5)\times10^4$~K.

 The optical depth of a given fragment is $\tau_{\rm f}= \kappa_{\rm f} \rho_{\rm f} R_{\rm f}$. 
 If $\tau_{\rm f}< 1$, a fragment is selected as a flare candidate otherwise it is ignored and it continues to contribute to the LFBOT light-curve. 
 Around the transparency time, $t_{\rm QN, tr}$, for example, when  $\rho_{\rm QN, tr}= M_{\rm QN}/4\pi R_{\rm QN, tr}^2\Delta R_{\rm QN, tr}\sim 10^{-16}$ g cm$^{-3}$, 
 we estimate $\tau_{\rm f, tr}\sim 0.1$ for fiducial parameter values.

 The radiation energy stored in a fragment of volume $V_{\rm f}$  is $E_{\rm f, rad.}\simeq aT_{\rm f}^4 V_{\rm f} \sim E_{\rm HS, SpD}(t_{\rm f}) \times (V_{\rm f}/V_{\rm QN, f})$
 since $E_{\rm HS, SpD} = aT_{\rm BB}(t_{\rm f})^4 V_{\rm QN, f}$; $V_{\rm QN, f}$ is the volume of the QN ejecta at $t_{\rm f}$ and $a$ the radiation constant. 
  When a fragment becomes optically thin ($\tau_{\rm f}\lesssim1$), it does \emph{not} radiate as an optically thick blackbody from its surface. Instead, the radiation energy stored throughout its volume at $t_{\rm f}$ is released on a timescale $\Delta t_{\rm f}\sim\tau_{\rm f}R_{\rm f}/c$, which for $\tau_{\rm f}\lesssim1$ approaches the light-crossing time $R_{\rm f}/c$.
 
  The duration of a flare is then 
 \begin{equation}
 \label{eq:dtf}
 \Delta t_{\rm f, pk}\sim 55.6\ {\rm mn}\times R_{\rm f, pk, 14}\ ,
 \end{equation}
 which is in tens of minutes for a fiducial peak value of $R_{\rm f, pk}=10^{14}$ cm.
 
 Here, $\Delta t_{\rm f,pk} = R_{\rm f,pk}/c$ denotes the characteristic (peak) flare duration associated with the Weibull scale $R_{\rm f,pk}$. For an individual fragment of radius $R_{\rm f}$, the flare duration is $\Delta t_{\rm f} = R_{\rm f}/c$, where $R_{\rm f}$ is sampled from the corresponding size distribution.

 The resulting luminosity is
\begin{align}
\label{eq:Lf}
L_{\rm f}
&= \frac{E_{\rm f,rad}}{\Delta t_{\rm f}}
\sim \frac{E_{\rm HS,SpD}(t_{\rm f})}{\Delta t_{\rm f}}\left(\frac{m_{\rm f}}{M_{\rm QN}}\right) \nonumber\\
&\sim L_{\rm LFBOT,pk}\left(\frac{t_{\rm QN,d}}{\Delta t_{\rm f}}\right)\left(\frac{m_{\rm f}}{M_{\rm QN}}\right),
\end{align}
where we used $\rho_{\rm f}\sim\rho_{\rm QN,f}$ and $E_{\rm HS, rot}\sim L_{\rm LFBOT,pk}t_{\rm QN,d}$. Since $(t_{\rm QN,d}/\Delta t_{\rm f})(m_{\rm f}/M_{\rm QN})\sim10^{-1}$--1, the fragment luminosity can reach values comparable to the LFBOT peak.

  The computed  flare luminosity is compared to those in AT2022tsd in Figure \ref{fig:LC-tsd}. Flares occurring during the LFBOT
   peak phase will be hard to distinguish from the LFBOT (as in AT2018cow; see \S~\ref{sec:other-LFBOTs}).

  The selection criterion $\tau_{\rm f} < 1$ restricts viable flare candidates to a specific region of parameter space. At $t_{\rm f}\sim t_{\rm QN,tr}\sim15$~days for example, the shell density is $\rho_{\rm f}\sim10^{-15}$~g~cm$^{-3}$, and a fragment with $R_{\rm f}\sim 10^{14}$~cm has $\tau_{\rm f}\sim\kappa_{\rm f}\rho_{\rm f}R_{\rm f}\sim 1.0 \kappa_{\rm f,1}$. Since fragment opacity is drawn from $1\lesssim\kappa_{\rm f}\lesssim100$~cm$^2$~g$^{-1}$, fragments in the \emph{low-$\kappa_{\rm f}$} tail of this distribution can simultaneously satisfy $\tau_{\rm f}<1$ and carry sufficient mass ($m_{\rm f}/M_{\rm QN}\sim10^{-4}$-$10^{-3}$) to produce flare luminosities approaching $L_{\rm LFBOT,pk}$ via Eq.~(\ref{eq:Lf}).  The most massive fragments sampled and shown in the left sub-panels in Figures \ref{fig:LC-tsd}, \ref{fig:LC-xnd-mrf} and \ref{fig:LC-cow}
  all correspond to lowest  values of the sampled $\kappa_{\rm f}$ despite forming early in time.
 The earlier a fragment forms, the denser it tends to be, making it less likely to be optically thin, since 
$\tau_{\rm f} = \kappa_{\rm f}\rho_{\rm f} R_{\rm f} \propto \frac{R_{\rm f}}{t_{\rm f}^3}$.
An exception occurs when the fragment’s radius, and therefore its mass, is relatively small, in which case any flaring would be negligible compared to the luminosity of the LFBOT.
 
 In general, a simultaneous fit to the LFBOT LC and to the optical flares can be obtained
   by  carefully choosing  the shape and scale parameters used for the WPD and WFTD.  
  We find that  a reasonable spread in radius ($k_{\rm radius}=1.5$) and formation time ($k_{\rm time}=1.5$)  reproduces
  AT2022tsd optical emission.  The case shown in Figure \ref{fig:LC-tsd} is for the parameters listed in Table \ref{table:parameter-fits}
  with $\alpha_{\rm f}=1.0$ which is when the peak fragment formation time coincides with the QN ejecta optical transparency.
    The left panels  show $t_{\rm f}, \Delta t_{\rm f}$ and $E_{\rm f, rad}$ (from top to bottom) of the resulting,  optically thin, fragments versus their mass $m_{\rm f}$ (in units of $10^{-3}M_{\rm QN}$). The x-axis shows $m_{\rm f}$, ordered from smallest to largest, to reflect the relationship $E_{\rm f, rad}\propto m_{\rm f}$.
    Figure \ref{fig:compae-to-Ho} compares the modeled flare duration and radiated energy to the observed ones. Best  agreement can be achieved
    when randomly varying the fragments opacity so that $1 < \kappa_{\rm f}\ ({\rm cm^2\ gm^{-1}}) < 100$ which is reasonable given the
  QN ejecta's heavy composition. 
  
  Because the flare durations are expected to follow a Weibull distribution (with $\Delta t_{\rm f} \propto R_{\rm f}$), and the formation times (specifically the intervals between flare arrival times) are expected to follow a WFTD, we performed a quantile-quantile (Q-Q) test \citep{wilk_1968} for the case of AT2022tsd to assess consistency between the observed and theoretical distributions.

In this analysis, the empirical quantiles correspond to the sorted observed values, while the theoretical quantiles were computed by applying the inverse cumulative distribution function (CDF) of the fitted model to uniformly spaced probabilities. Despite the limited sample size, Figure~\ref{fig:Q-Qplot} shows that the theoretical quantiles broadly align with the 45$^\circ$ reference line, indicating tentative agreement. However, a more robust statistical analysis is required to confirm this result.

\subsection{X-ray: the HS spin-down}
\label{sec:xray}

The resulting X-ray luminosity is proportional to the HS spin-down power,
$L_{\rm x}\sim\eta_{\rm x}L_{\rm HS,SpD}$, with conversion efficiency
$\eta_{\rm x}\lesssim1$; we adopt $\eta_{\rm x}\sim1$ for simplicity.
For $t>t_{\rm HS,SpD}$, the spin-down luminosity scales as
$L_{\rm HS,SpD}\propto B_{\rm HS,15}^{-2}t^{-2}$,
becoming independent of the HS rotation period, i.e.,
\begin{equation}
L_{\rm x} \sim 2\times10^{45}\ {\rm erg\ s}^{-1}\ B_{\rm HS,15}^{-2}\ t_{\rm day}^{-2},
\end{equation}
where $t_{\rm day}$ is time in days. This emission escapes preferentially
through low-optical-depth gaps (associated with transparent fragments), appearing intermittent
and discontinuous.  As shown in Figure~\ref{fig:LC-tsd}, this relation help reproduce the observed X-ray luminosity of AT2022tsd, implying that 
a fraction of the spin-down energy is emitted directly as X-rays due to the fragmented nature of the QN ejecta.

\subsection{Radio: the QN shock}
\label{sec:radio}

Radio emission arises from electrons accelerated at the shock front as the non-relativistic QN ejecta expands and interacts with the surrounding medium. These mildly relativistic electrons follow a power-law energy distribution with slope $2 < p < 3$. The spectral luminosity at an observed frequency $\nu_{\rm obs}$ is given by (e.g., \citealt{nakar_2011,granot_2006,sironi_2013}):

\begin{align}
    L_{\nu_{\rm obs}} (t) &=  L_{\rm pk, dec} \left(\frac{\nu_{\rm pk, dec}}{\nu_{\rm obs}} \right)^{(p-1)/2} \times \\\nonumber
    &\times
    \begin{cases}
        \left(\frac{t}{t_{\rm dec}}\right)^{\frac{3(7-5p)}{10}}, & t \geq t_{\rm dec} \\
        \left(\frac{t}{t_{\rm dec}}\right)^3, & t < t_{\rm dec}
    \end{cases}
\end{align}
Here, $\nu_{\rm pk, dec} \sim 78.4~{\rm GHz}~g(p)^2 \epsilon_{\rm e, -1}^2 \epsilon_{\rm B, -2}^{1/2} E_{\rm QN, 50} n_{\rm amb, 6}^{1/2} t_{\rm dec, day}^{-3}$ is the peak frequency at the deceleration time $t_{\rm dec}$ (in days), and $L_{\rm pk, dec} \sim 1.7 \times 10^{33}~{\rm erg~s^{-1}~Hz^{-1}}~\epsilon_{\rm B, -2}^{1/2} E_{\rm QN, 50}^{4/5} n_{\rm amb, 6}^{7/10} t_{\rm dec, day}^{3/5}$ is the corresponding peak luminosity. The parameters $\epsilon_{\rm e}$ and $\epsilon_{\rm B}$ are the fractions of the internal energy in electrons and magnetic fields, respectively, and $g(p) = (p-2)/(p-1)$.

Since the peak synchrotron frequency $\nu_{\rm pk, dec} \sim \nu_{\rm m}$ (corresponding to the minimal Lorentz factor $\gamma_{\rm m}$) and synchrotron self-absorption frequencies are below $\nu_{\rm obs}$, the deceleration time also marks the radio peak. 

For a given ambient density $n_{\rm amb}$, which sets $t_{\rm dec}$, we fit the radio emission by adjusting $\epsilon_{\rm e}$, $\epsilon_{\rm B}$, and $p$, as listed in Table~\ref{table:parameter-fits}. This model reproduces the observed radio light curve of AT2022tsd, as shown in Figure~\ref{fig:LC-tsd}.

%
 %
\begin{figure*}[t!]
\begin{center}
\includegraphics[scale=0.4]{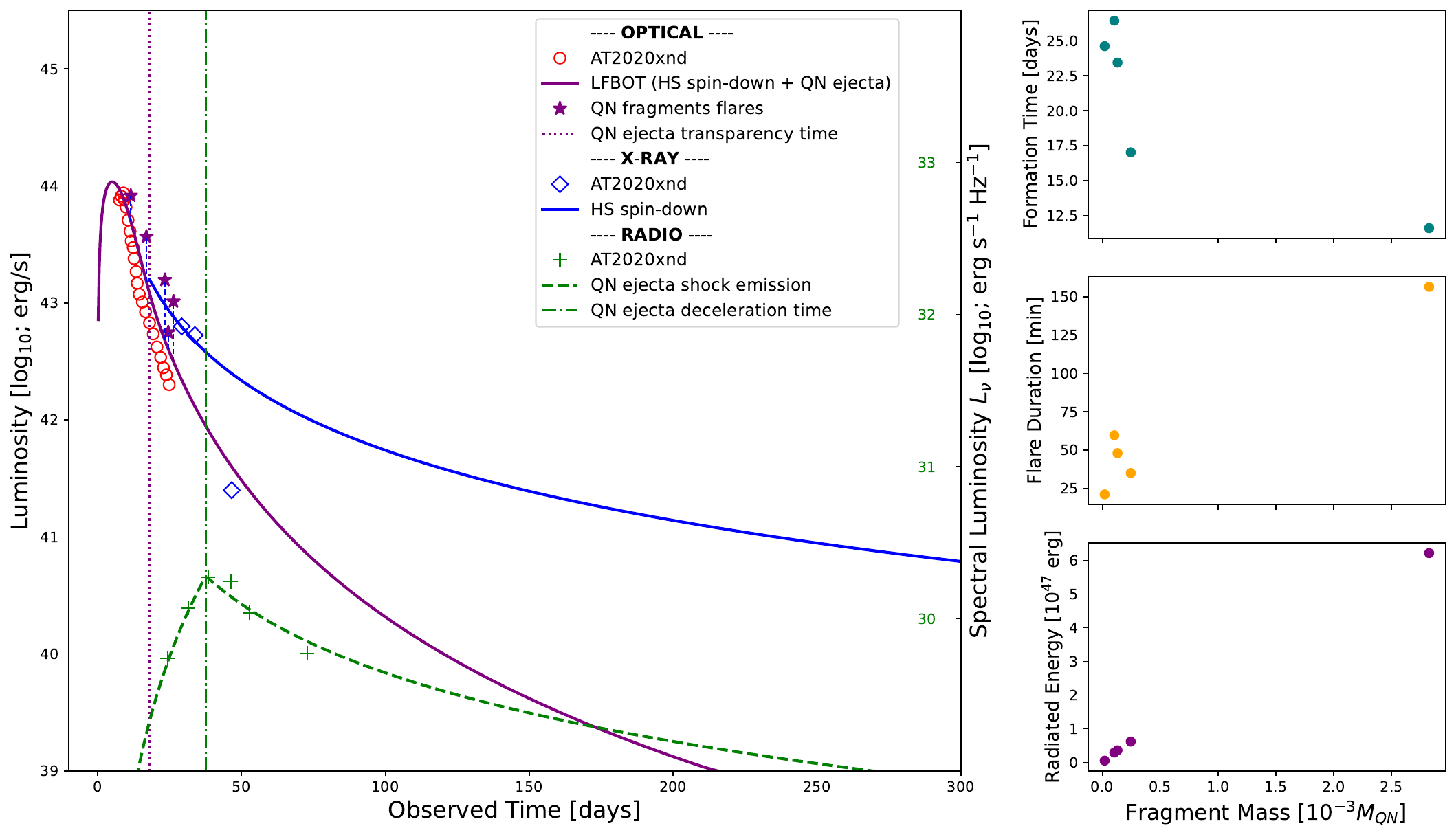}
\includegraphics[scale=0.4]{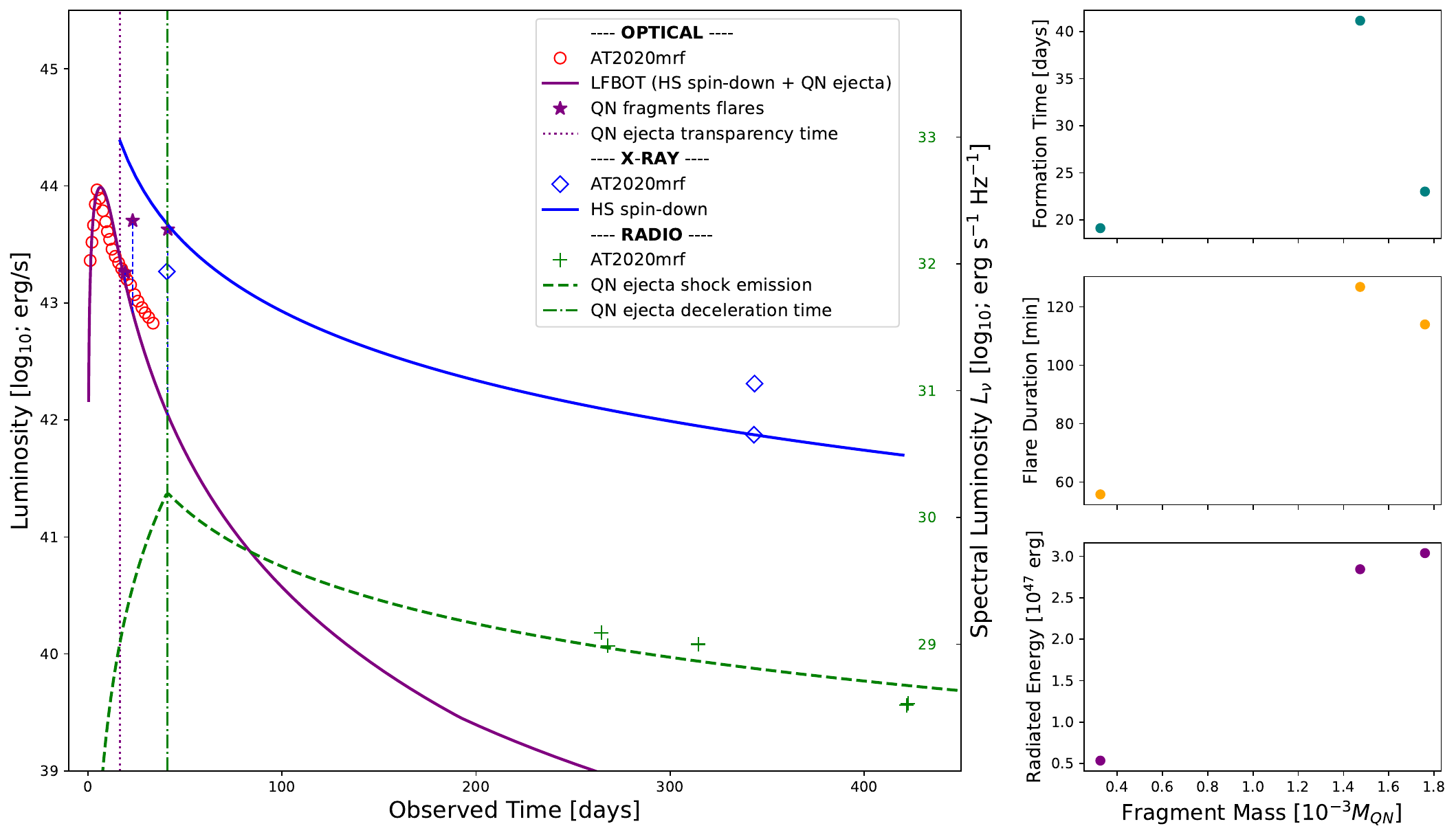}
\caption{Same as in Figure \ref{fig:LC-tsd}, but for AT2020xnd (top panel) and AT2020mrf (lower panel). 
The radio data for AT2020mrf  is from \citet{yao_2022}}
\label{fig:LC-xnd-mrf}
\end{center}
\end{figure*}

\begin{figure*}[t!]
\begin{center}
\includegraphics[scale=0.4]{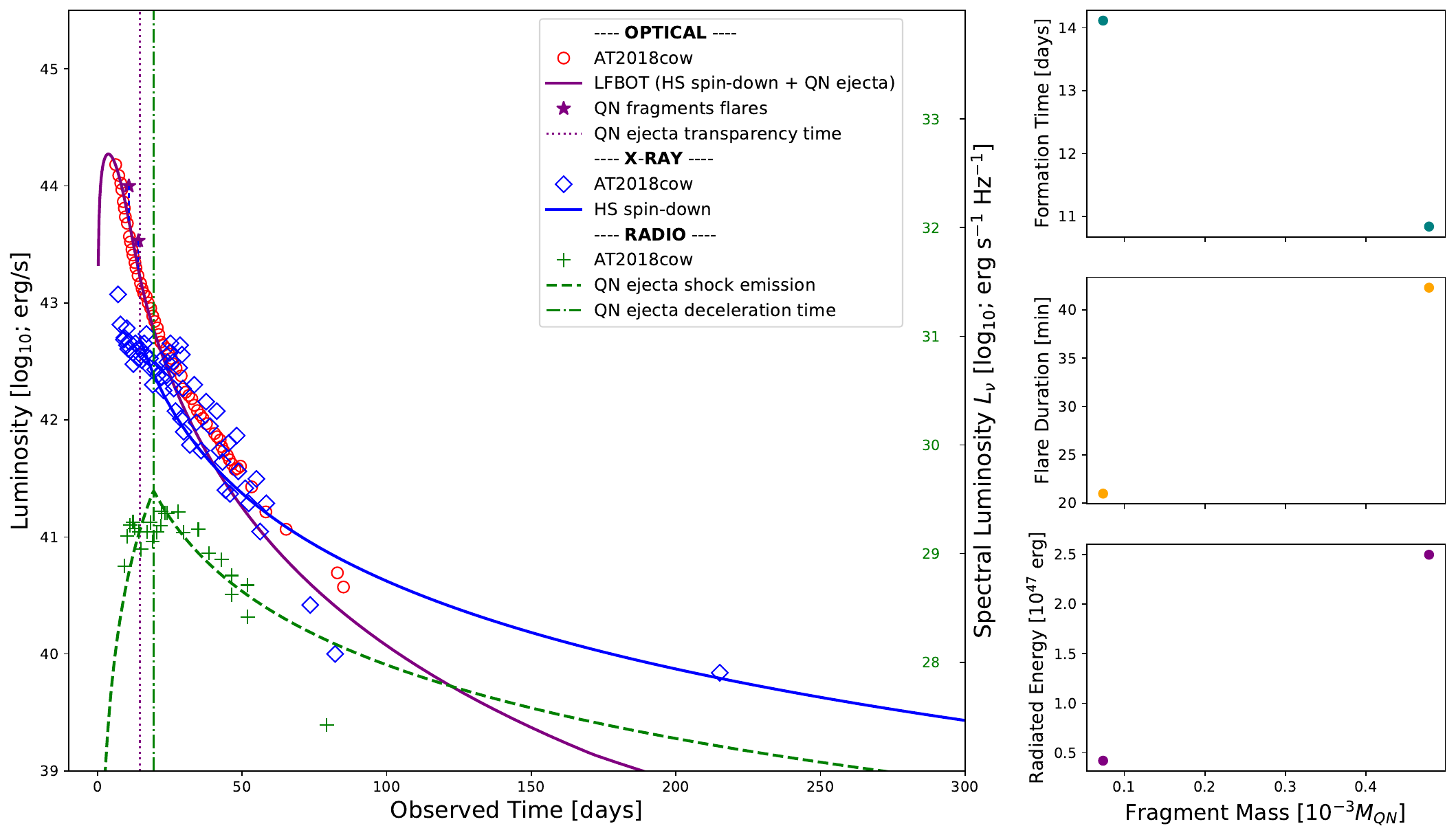}
\caption{Same as in Figure \ref{fig:LC-tsd}, but for AT2018cow.}
\label{fig:LC-cow}
\end{center}
\end{figure*}

%
 %

\subsection{AT2020xnd, AT2020mrf  and AT2018cow}
\label{sec:other-LFBOTs}

Fits to AT2020xnd and AT2020mrf  can be obtained for parameters similar to those used for AT2022tsd as listed in
Table \ref{table:parameter-fits}. To obtain
no or fewer optically thin fragments (i.e. flares), we simply consider a scenario where fragmentation starts earlier than in the case 
of AT2022tsd; e.g. for $\alpha_{\rm f}=t_{\rm break}/t_{\rm QN, tr}=0.3$ as portrayed in the cases shown in Figure \ref{fig:LC-xnd-mrf}.  

We explained earlier that when fragmentation occurs before or during the  LFBOT LC 
   peak phase, the flares will be hard to distinguish from the LFBOT (see Eq. (\ref{eq:Lf})).
   Writing $t_{\rm f, pk}=\alpha_{\rm f}t_{\rm QN, tr}< t_{\rm QN, d}$ gives
   $\alpha_{\rm f} < 0.4 E_{\rm QN, 50}^{1/2}/M_{\rm QN, -2}^{1/4}E_{\rm SN}$. 
   Furthermore, the early a fragment forms the denser it is and the less probable  to be born optically thin given that  $\tau_{\rm f} =\kappa_{\rm f}\rho_{\rm f} R_{\rm f}\propto R_{\rm f}/t_{\rm f}^3$. Unless the fragment radius (and thus its mass) is relatively small in which case flaring will be dwarfed by the LFBOT luminosity.
    
 The Fits to the observed X-ray luminosity in AT2020xnd and AT2020mrf, also shown in Figure \ref{fig:LC-xnd-mrf}, show
a good agreement with  AT2020mrf while in the case of AT2020xnd the last 
data point is too low.  Both require  an efficient  conversion of HS SpD energy
to X-rays with $L_{\rm x}\sim L_{\rm HS, SpD}$.  The modelled radio emission also agrees with observations with an ambient density very similar to that of AT2022tsd. 

Fits to the optical light-curve of AT2018cow required a slightly faster spinning HS and 
a relatively higher $n_{\rm amb}$ value compared to the other three LFBOTs when fitting the radio emission (see Table \ref{table:parameter-fits}).
  Most runs yield no flares when $\alpha_{\rm f}< 0.3$ with a  few occurring during the LFBOT peak diffusion phase when $\alpha_{\rm f}=0.3$.
   Their luminosity is  similar to $L_{\rm LFBOT, pk}$ and would be hard to differentiate  from the main LC;  see Figure \ref{fig:LC-cow}. The 
earlier deceleration phase from the higher $n_{\rm amb}$ (Eq. (\ref{eq:tdec}))  is consistent with this picture since 
  fragmentation would occur while the ejecta is still dense and the LFBOT still at its peak phase.  
 It also required the highest value of $p$ which may mean that the early fragmentation of the ejecta creates less favorable electron acceleration conditions.
  
  The observed X-ray emission can be fit with a decay profile steeper than $t^{-2}$, with a $t^{-2.5}$ slope providing a decent fit to most of the data. However, AT2018cow appears to require an additional emission component (see Figure~\ref{fig:LC-cow}), suggesting more complex -- or possibly different -- behavior than a simple power-law decay. It is important to note that the X-ray emission expected from the same electron population responsible for the radio synchrotron emission is too weak to account for this excess. One possible contributor could be magnetar-like activity from the HS, though a detailed exploration of this scenario is beyond the scope of this paper.

\section{Discussion and Limitations }
\label{sec:discussion}

There are  unusual properties of LFBOTs worth discussing:

(i) \textit{The Hot, Blue, Featureless Spectrum}: The effective blackbody
temperature of the QN ejecta (Eq.~(\ref{eq:TBB})) may account for the hot, blue continuum observed at early times in LFBOTs.
 The temperature  evolves as
$T_{\rm BB}\propto t^{-1/4}$ for $t<t_{\rm QN,d}$ and as
$T_{\rm BB}\propto t^{-1/2}$ for $t>t_{\rm QN,d}$ which can be tested against observed spectra of LFBOTs.
 For example, for AT2022tsd parameters ($P_{\rm HS}=8.5$~ms,
$B_{\rm HS}=6\times10^{14}$~G), Eq.~(\ref{eq:TBB}) gives
$T_{\rm BB}\sim3.8\times10^4$~K at $t=1$~day and $\sim1.5\times10^4$~K
at $t=15$~days; this may be compared to the observed featureless blue continuum.

The combination of the QN ejecta's low mass and high velocity should yield 
strong Doppler broadening and line blending, which may wash out
potential absorption features at early times. Rapid optical flares are
expected to coincide with a spectral shift from a purely thermal BB to
a composite spectrum that includes non-thermal contributions, though the
underlying BB component remains at least partially discernible;  this
spectral transition represents a testable prediction of the model.
Hydrogen and helium lines, observed in some LFBOTs at late times, are
unlikely to be intrinsic to the neutron-rich QN ejecta. They likely arise
from mixing between the QN ejecta and the surrounding ambient medium,
becoming visible only after significant deceleration and interaction with
swept-up material.

\textit{(ii) Offsets and Long-Period Magnetars.}
For sufficiently long delays, the HS transition (and thus QCD magnetar formation and the associated LFBOT) can occur far from the original star-forming region. In such cases, the transient may appear significantly offset from its host galaxy, or even apparently hostless in extreme scenarios. As a plausibility example, this would be reminiscent of events such as AT2023fhn \citep{chrimes_2024a}.

\textit{(iii) Long-period QCD magnetars.} If the NS reaches the transition as a slow rotator, the phase transition produces a slowly rotating QCD magnetar with no or very weak LFBOT.
 These could manifest as a long-period magnetar with an old kinematic age and no associated luminous transient. 
ASKAP~J1832$-$0911 \citep{wang_2024}, a long-period radio transient with magnetar-like activity and no detected associated transient, 
may be considered as a possible analogue of such a system.
   
 (iv) {\it LFBOT rate estimate}:  We adopt a log-normal distribution for the NS birth magnetic field, characterized by a mean of $\mu_{\log B_{\rm NS}} = 12.5$ and a standard deviation of $\sigma_{\log B_{\rm NS}} = 0.5$. For the birth spin periods, we use a normal distribution with a mean of $\mu_{P_{\rm NS}} = 300$ ms and a standard deviation of $\sigma_{P_{\rm NS}} = 150$ ms (e.g., \citealt{faucher_2006}).

The maximum LFBOT rate is then estimated as:
\begin{align}
\label{eq:GRB-rate}
r_{\rm LFBOT,\ max} &= \left[ \frac{ \int_{10}^{13} \exp\left(-\frac{(\log{B_{\rm NS}} - \mu_{\log{B_{\rm NS}}})^2}{2\sigma_{\log{B_{\rm NS}}}^2}\right)\, d\log B_{\rm NS}}{\int_{10}^{15} \exp\left(-\frac{(\log{B_{\rm NS}} - \mu_{\log{B_{\rm NS}}})^2}{2\sigma_{\log{B_{\rm NS}}}^2}\right)\, d\log B_{\rm NS}} \right] \times \nonumber \\
&\quad \left[ \frac{ \int_{1.5}^{50} \exp\left(-\frac{(P_{\rm NS} - \mu_{P_{\rm NS}})^2}{2\sigma_{P_{\rm NS}}^2}\right)\, dP_{\rm NS}}{\int_{1.5}^{\infty} \exp\left(-\frac{(P_{\rm NS} - \mu_{P_{\rm NS}})^2}{2\sigma_{P_{\rm NS}}^2}\right)\, dP_{\rm NS}} \right] \times \nonumber \\
&\quad r_{M_{\rm prog}} \times r_{\rm CCSN},
\end{align}
where $r_{M_{\rm prog}}$ is the fraction of CCSNe that give birth to NSs, and $r_{\rm CCSN}$ is the volumetric CCSN rate. 
Naturally, only a subset of these NSs are expected to undergo conversion (\citealt{ouyed_2025,ouyed_2026}).

Evaluating the above with representative values, we obtain
\begin{equation}
r_{\rm LFBOT,\ max} \sim 0.8 \times 0.02 \times 0.8 \times r_{\rm CCSN} \sim 0.01\, r_{\rm CCSN},
\end{equation}
 comparable to the estimated LFBOT rate of $\lesssim 0.1\%$ of the local CCSN rate (\citealt{coppejans_2020, ho_2023b}).

(vi) \textit{Kilonova-like emission and LFBOTs}: The QN ejecta, as a plausible site of $r$-process nucleosynthesis \citep{jaikumar_2007}, may exhibit kilonova-like emission (\citealt{paczynsky_1980, metzger_2010}). However, unlike conventional kilonovae associated with compact object mergers, QNe can occur across a wide range of environments and are not necessarily tied to specific metallicity conditions. 
In particular, QNe occurring in isolation (and their associated LFBOT signatures) could contribute to heavy-element enrichment in environments and on timescales distinct from those of compact object mergers.  A unique prediction of our model is the eventual associations of LFBOTs with kilonovae in environments that are inconsistent with compact mergers.

(v) \textit{Physical distinction between QN and merger ejecta}: 
In NS mergers, tidal forces and shocks strip the crust from the outside inward, rapidly heating it into a high-entropy, fully fluid plasma. Any initial crystalline structure or mechanical coherence is erased, and density inhomogeneities are efficiently smoothed out.
In contrast, in a QN the outer layers are expelled from the inside out (radial expansion with no preferred directional ejection) by the energy released during the core quark--hadron phase transition \citep{keranen_2005,ouyed_leahy_2009}. The crust is accelerated mechanically before being strongly shock-heated, allowing it to retain short-lived density inhomogeneities and partial coherence during early expansion. This provides a physically motivated basis for differences in geometry, velocity structure, and early radiative behaviour relative to merger ejecta.

 Notably, the lack of flares in  LFBOTs may indicate that strong fragmentation is not ubiquitous.  One possible explanation, is that 
 most of the optically thin fragments occur during the diffusion phase (i.e. $\alpha_{\rm f}=t_{\rm break}/t_{\rm QN, tr}\le 0.3$; see Table \ref{table:parameter-fits})
 in which case they will be dwarfed by the LFBOT emission proper.   In that case, AT2022tsd may represent a special realization of the outflow conditions rather than the norm. 
We emphasize, however, that the presence of clumping remains a working assumption of our model; alternative mechanisms may produce ejecta patchiness, and a detailed treatment of the fragmentation process is beyond the scope of this study.

\section{Conclusion}
\label{sec:conclusion}

The delayed conversion of a NS into a highly magnetized HS (a QCD-magnetar
in our model) ejects $\sim10^{-2}\,M_\odot$ of neutron-rich outer layers
with kinetic energy $\sim10^{50}$~erg (the QN ejecta).
The HS acquires its strong field from the quark phase (which remains to be determined) in its deconfined core 
independently of the NS birth spin. If the HS inherits rapid millisecond rotation from its progenitor NS 
or from ongoing accretion from a companion, its spin-down energy powers the QN ejecta, producing
a transient with peak luminosity, rise and decay times of a few days consistent with LFBOTs.

 For this transition to occur, the NS must first reach the critical mass $M_{\rm NS,c}\sim 1.8\,M_\odot$
for quark deconfinement to set in. This can happen in two ways: either the NS
is born sufficiently massive following a CCSN, in which case the QN may follow
while still embedded in dense SN ejecta yielding a SLSN rather than an
LFBOT \citep{ouyed_2026}. The NS may also grow toward $M_{\rm NS,c}$ through
accretion from a binary companion, a process that should result in an LFBOT in relative isolation. 
The resulting LFBOT is directly observable and represents the distinctive
optical signature of the QN event in the binary accretion channel (or in isolation in general).

A working hypothesis of our model is the fragmentation of the QN ejecta  because the QN ejecta is launched from
the inside out, in contrast to the outside-in tidal disruption that
characterizes NS mergers. The ejecta is accelerated without strong shock
heating, allowing it to remain cold and dense during early expansion. These
properties, we speculate, enable the development of transient inhomogeneities within the
ejecta. Under suitable conditions, the ejecta fragments, producing optically
thick structures that become transparent (during the LFBOT phase) on timescales of tens of minutes and
release the trapped LFBOT radiation. The trapped energy is radiation energy
stored throughout the fragment volume, $E_{\rm f,rad}=aT_{\rm f}^4V_{\rm f}$;
when the fragment optical depth falls below unity, this energy escapes on the
light-crossing timescale which is of order of minutes, powered by the HS spin-down injection that maintains
the radiation temperature. This gives rise to optical flares with
luminosities comparable to the LFBOT peak.

X-ray emission originates from the relativistic wind
driven by HS spin-down power, with synchrotron and inverse Compton emission
from the shocked wind producing non-thermal X-rays that leak through
low-optical-depth regions in the fragmented QN ejecta. Radio emission arises from
synchrotron radiation as the ejecta interacts with the ambient medium. The
model reproduces key features of AT2022tsd as well as other LFBOTs, including
AT2020xnd, AT2020mrf, and AT2018cow, with the presence or absence of prominent
flares determined by the timing of fragmentation relative to the peak emission
phase.

The neutron-rich composition of the QN ejecta, together with the possibility
of QNe occurring in isolated environments, implies that this channel contributes
to heavy-element enrichment in a way that is fundamentally distinct from compact
object mergers (see \citealt{ouyed_2009}).  We therefore predict the existence of LFBOTs accompanied
by kilonova-like emission -- temporally and spatially coincident -- in
environments that do not host neutron star mergers, representing a smoking-gun
signature of the QN channel.

\end{document}